\documentstyle[12pt]{article}

\makeatletter
\def\section{\@startsection{section}{1}{\z@}{-3.5ex plus -1ex minus -.2ex}
{2.3ex plus .2ex}{\large\bf}}
\makeatother
 
\makeatletter
\@addtoreset{equation}{section}

\makeatother

\def\lb{\lbrack}
\def\rb{\rbrack}

 \def\Slash#1{
  \begin{picture}(5,6)(0,0)
  \put(-.7,-1.2){\line(5,6)6}
  \end{picture}
  \kern-.8em#1}
 \def\slash#1{
  \begin{picture}(5,6)(0,0)
  \put(-1.5,-1.7){\line(5,6)5}
  \end{picture}
  \kern-.8em#1}

\def\Sn{\Slash \nabla}

\def\gg5{\gamma_5}
\def\hg5{\hat{\gamma}_5}
\def\g4{\gamma_4}

\def\O{{\cal O}}

\def\wD{\widetilde{D}}
\def\wnabla{\widetilde{\nabla}}

\def\Qlatmr1{Q_{lat}^{(m=r=1)}}

\def\be{\begin{eqnarray}}
\def\ee{\end{eqnarray}}

\def\bx{{\bf x}}
\def\bp{{\bf p}}

\parskip=\medskipamount

\begin{document}

\input epsf

\vspace{8mm}

\begin{center}
 
{\Large \bf On the fourth root prescription for dynamical staggered fermions}
\\

\vspace{0.3ex}

\vspace{12mm}

{\large David H. Adams}

\vspace{4mm}

Instituut-Lorentz for Theoretical Physics, Leiden University, \\
Niels Bohrweg 2, NL-2333 CA Leiden, The Netherlands\\

and \\

National Center for Theoretical Sciences, National Taiwan University,\\ 
Taipei 106, Taiwan R.O.C.\\

and \\

Physics Dept., Florida International University, \\
11200 SW 8th St., Miami, FL 33199, U.S.A.
\footnote{Current address}\\

\vspace{1ex}

email: adams@lorentz.leidenuniv.nl

\end{center}

\begin{abstract}

With the aim of resolving theoretical issues associated with the fourth
root prescription for dynamical staggered fermions in Lattice QCD simulations, we consider
the problem of finding a viable lattice Dirac operator $D$ such that 
$(detD_{staggered})^{1/4}=detD$. Working in the flavour field representation 
we show that in the free field case there is a simple and natural candidate $D$ satisfying this 
relation, and we show that it has acceptable locality behaviour: exponentially local 
with localisation range vanishing $\sim\sqrt{a/m}$ for lattice spacing $a\to0$.
Prospects for the interacting case are also discussed, although we do not solve this case here.

\vspace{3cm}
\end{abstract}

\pagebreak

\section{Introduction}

The development in recent years of an improved staggered fermion formulation \cite{improved} 
has made unquenched numerical Lattice QCD simulations possible at realistically small quark masses.
The resulting impressive agreement between the calculated parameters of QCD phenomenology 
and their experimental values \cite{Davies(PRL),Bernard} (along with predictions for quantities
not yet measured experimentally \cite{Allison})
indicates that the long-time dream of being able to
do high-precision Lattice QCD calculations is now becoming reality. 
However, the advantageous
properties of staggered fermions for numerical implementation are currently offset by unresolved
problematic issues at the conceptual/theoretical level. In particular, there is concern
\cite{Jansen(L2003),DeGrand,Neu} about the use of the fourth root of the staggered fermion
determinant to represent the fermion determinant of a single dynamical (sea) quark. A number
of works have appeared recently addressing this concern via theoretical considerations
\cite{DA,Jansen(NPB),Jansen(L2004),Hart,Neu(root),Peardon,Shamir,Giedt}, 
derivations of predictions that
can be used to test its viability \cite{Sharpe}, and various numerical investigations
\cite{Durr,Follana,Woloshyn}. 
The present paper is intended as another theoretical contribution in this direction.

A staggered fermion is a lattice formulation of four continuum fermion flavours, nowadays called 
``tastes'' (to distinguish them from the actual quark flavours). The fermion determinant for a 
single quark flavour in this framework is represented by a rooted determinant
$det(D_{staggered})^{1/4}$. While this formally goes over to the determinant for a single quark 
flavour in the continuum limit, the concern regarding this prescription is that it does not fit
in an obvious way into the framework of local lattice field theory at non-zero lattice spacing.
The lattice model might therefore not be in the right universality class to reproduce QCD.
This raises the question of whether the dynamical staggered fermion formulation is a first 
principles approach to QCD or simply a phenomenological model which describes QCD very well in
a certain regime.

One way to establish that the universality class is the right one would be to show
that there is a viable (and in particular, local) single-flavour lattice Dirac operator $D$ 
such that \cite{Jansen(L2003)}
\be
(detD_{staggered})^{1/4}=detD
\label{1.1}
\ee
This would imply equivalence between the dynamical staggered fermion formulation and the manifestly
local formulation with sea quarks described by $D$. (To avoid unitarity issues, $D$, with suitably 
adjusted bare mass, should then also be used as the Dirac operator for the valence quarks.)
We will refer to (\ref{1.1}) as the Staggered Determinant Relation (SDR) in the following.

The most direct attempt at a solution to the SDR is simply to take $D=(D_{staggered})^{1/4}$.
This is essentially the approach taken by Jansen and collaborators in Ref.\cite{Jansen(NPB)}.
More precisely, they considered the operator
\be
M=\Big((D_{staggered})^{\dagger}D_{staggered}\Big)^{1/2}\Big|_{even\ sites}
\label{1.2}
\ee
for which $detM=det(D_{staggered})^{1/2}$ since $(D_{staggered})^{\dagger}D_{staggered}$ couples
lattice sites by even-to-even and odd-to-odd. Thus $M$ is a candidate operator for $D$ in the
case where $1/4\to1/2$ in (\ref{1.1}), i.e. the case of two degenerate
quark flavours. However, this operator was found to have unacceptable locality behaviour:
it is exponentially local (for bare mass $m>0$), but the localisation range is $\sim m^{-1}$ and
thus fails to vanish in the limit of vanishing lattice spacing, $a\to0$ \cite{Jansen(NPB)}.

This negative result of Ref.\cite{Jansen(NPB)} is unsurprising, since the operator (\ref{1.2}) 
does not take account of the staggered fermion taste structure. The staggered fermion action can 
be viewed as consisting of
naive fermion actions for four fermion species (the tastes), together with terms
that couple these, with the latter formally vanishing for $a\to0$. This suggests that, in 
attempting to find a local, single-flavour lattice Dirac operator satisfying the SDR, one should 
consider operators of the form 
\be 
D=\Sn+W+m
\label{1.3}
\ee
where $\Sn$ is the (massless) naive lattice Dirac operator, and $W$ is a term which formally 
vanishes for $a\to0$ and whose role is to take account of the taste-mixing in $D_{staggered}$.
In this paper, working in the flavour field representation of staggered fermions \cite{Kluberg},
we show that, in the {\em free field case}, there is a natural candidate $D$ of the form 
(\ref{1.3}) which satisfies the SDR. The $W$ operator, although formally vanishing for $a\to0$,
turns out to involve a square root, so its locality status is not immediately clear. 
We show, however, that the operator does have acceptable locality behaviour: exponentially
local, with localisation range vanishing $\sim\sqrt{a/m}$ for $a\to0$.

Our operator can be gauged (i.e. coupled to the link variables) in a variety of ways. However,
for reasons which we will discuss later, it is most unlikely that a gauging of this operator 
exists such that the SDR continues to hold in the interacting case.
Our operator should therefore be regarded as a prototype, i.e. a first step on a path to
constructing more sophisticated operators which have a chance of satisfying the SDR in the
full interacting case.

Of course, there is no a priori guarantee that a viable lattice Dirac operator satisfying 
the SDR actually exists in the interacting case, so other approaches should also be considered.
One possibility is the following: If there is a single-flavour $D$ such that the effect
of including the determinant ratio $det(D_{staggered})^{1/4}/detD$ in the Lattice QCD functional
integral is simply to renormalise the bare coupling constant (just as dynamical heavy quarks do
\cite{Hasenfratz-DeGrand}), then representing the sea quark determinant by 
$det(D_{staggered})^{1/4}$ is equivalent to representing it by $detD$ together with a shift in the 
bare coupling. Since the latter description is manifestly local, this is another way in which the
locality issue could be positively resolved. The prospects for this, and the properties that such
a $D$ would be expected to have, are also discussed in some detail in this paper. 
The problem of finding such a $D$ is seen to be essentially equivalent to the problem of finding a
solution to a generalised version of the SDR.

The paper is organised as follows.
After a general discussion of the problem of finding viable solutions to the SDR, we arrive at 
our free field candidate $D$ in \S2. In \S3 we prove that this
operator has the good locality behaviour mentioned earlier. Our argument is entirely analytic and 
the techniques are of a generally applicable nature; we also apply them to give a new derivation 
of the negative locality result for the operator considered previously by Jansen and collaborators 
\cite{Jansen(NPB)}. (Their argument in the free field case had a numerical as well as analytic 
component.) We conclude in \S4 with a discussion of the issues and prospects for the interacting 
case.

\section{Considerations for finding a candidate $D$}

For concreteness we specialise to 4 spacetime dimensions in this section (everything
generalises straightforwardly to arbitrary even dimensions). The usual staggered fermion action,
obtained via spin-diagonalisation of the the naive action \cite{Smit}, is 
$S_{staggered}=a^4\sum_x\bar{\psi}(x)D_{st}\psi(x)$ where the staggered Dirac operator
is given by
\be
D_{st}=\eta^{\mu}\frac{1}{a}\nabla_{\mu}+m
\label{2.1}
\ee
with $\eta^{\mu}(x)=(-1)^{(x_1+\dots+x_{\mu-1})/a}$ and 
$\nabla_{\mu}=\frac{1}{2}(\nabla_{\mu}^++\nabla_{\mu}^-)$ the symmetrised 
gauge-covariant difference operator. The usual flavour (i.e. taste) identification comes about
by considering the free field propagator: it has 4 poles, and the momentum space Brillouin zone
is divided into 4 subregions, each containing a single pole, with the momenta in each of these 
subregions being interpreted as the momenta of different fermion tastes. 
An alternative, and conceptually more appealing way to identify the tastes is provided by 
the flavour (taste) field representation of the staggered fermion action derived in \cite{Kluberg}. 
In this representation the taste fields are manifest from the beginning in the fermion action.
The taste fields live on the blocked lattice (spacing=$2a$), whereas the lattice paths and 
link variables which specify the gauging of the action are those of the original lattice.
The action in general gauge background does not have a simple expression in this setting
though, making it more difficult to work with in practise. However, in the free field case the 
action does have a simple expression. Denoting the staggered Dirac operator in the taste field 
representation by $D_{stt}$, it can be written in the free field case as
\cite{Kluberg}\footnote{The free field version of the staggered Dirac operator in the taste
representation also arises from a first principles approach to constructing the Dirac operator
on the lattice \cite{Rossi-Wolff}.}
\be
D_{stt}^{free}=(\gamma^{\mu}\otimes{\bf 1})\frac{1}{2a}\nabla_{\mu}
+i(\gamma_5\otimes\Gamma^{\nu})\frac{1}{2(2a)}\Delta_{\nu}+m
\label{2.2}
\ee
where now the (free field) difference operators are on the blocked lattice;
$\Delta_{\nu}=\nabla_{\nu}^--\nabla_{\nu}^+$ (so that $\Delta\!=\!\sum_{\nu}\Delta_{\nu}$
is $(2a)^2$ times the blocked lattice Laplace operator), and $\{\Gamma^{\nu}\}$ 
is a hermitian representation
of the Dirac $\gamma$-algebra on taste space ${\bf C}^4$.\footnote{The part 
$i\gamma_5\otimes\Gamma^{\nu}$ in (\ref{2.2}) is usually written as 
$\gamma_5\otimes\tau_5\tau^{\nu}$ where $\{\tau^{\nu}\}$ is a hermitian representation of the 
Dirac $\gamma$-algebra on taste space. Note that $\Gamma^{\nu}=-i\tau_5\tau^{\nu}$ defines 
another (equivalent) hermitian representation of this algebra.}

The importance of taking account of taste structure when attempting to find a solution $D$ to
the SDR can now be seen in the free field case as follows.
The $\gamma$-matrix representation $\{\Gamma^{\nu}\}$ in (\ref{2.2}) can be chosen such that the
diagonal matrix elements in each of the $\Gamma^{\nu}$'s all vanish. Then the taste-mixing
terms in the free field lagrangian correspond to the terms with $\Gamma^{\nu}$'s in the free field 
Dirac operator (\ref{2.2}). Therefore, if the taste-mixing terms are ``turned off'' 
the free field Dirac operator reduces to $\Sn\otimes{\bf 1}+m$, where 
$\Sn=\gamma^{\mu}\frac{1}{2a}\nabla_{\mu}$ is the (massless) free field naive Dirac operator and 
${\bf 1}$ is the identity matrix on taste space. Consequently,
\be
(detD_{st}^{free})^{1/2}\;\to\;det\left({\Sn+m \atop 0}\ {0 \atop \Sn+m}\right)\quad,\quad
(detD_{st}^{free})^{1/4}\;\to\;det(\Sn+m)
\label{2.3}
\ee
or, alternatively,
\be
(detD_{st}^{free})^{1/2}&\to&det\left({\sqrt{(\Sn+m)^{\dagger}(\Sn+m)} \atop 0}\
{0 \atop \sqrt{(\Sn+m)^{\dagger}(\Sn+m)}}\right) \nonumber \\
(detD_{st}^{free})^{1/4}&\to&det\left(\sqrt{(\Sn+m)^{\dagger}(\Sn+m)}\;\right)\,.
\label{2.4}
\ee
In the former case the fractional powers of $detD_{st}^{free}$ become determinants of 
ultra-local lattice Dirac operators, 
while in the latter case they become determinants of operators which cannot be 
expected to have good locality properties. If we now imagine turning back on the taste-mixing
terms, there is reason to hope that there will be corresponding deformations of 
$\left({\Sn+m \atop 0}\;{0 \atop \Sn+m}\right)$ or
$\Sn+m$ into some two-taste lattice Dirac operator 
$\wD$ or single-taste $D$, respectively, which continues to have good locality behaviour, 
such that $(detD_{st}^{free})^{1/2}=det\wD$ and $(detD_{st}^{free})^{1/4}=detD$. 
On the other hand, if a solution $D$ to the SDR, or a 
solution $\wD$ to the version of the SDR with with fractional power $1/2$ of the staggered
fermion determinant, has been constructed ``blindly'' without taking account of the taste structure 
of the staggered fermion formulation, it can happen that when taste-mixing terms
are turned off in the free field case the scenario (\ref{2.4}) arises; then
it is to be expected that the $D$ or $\wD$ have bad locality behaviour. In fact this is essentially
the situation for the solution $\wD=M$ considered in \cite{Jansen(NPB)}, and the negative locality
result found there is therefore unsurprising. However, the possibility (\ref{2.3}) gives hope
of doing better than this, at least in the free field case.\footnote{The interacting case is
more difficult, since the taste field representation of the staggered Dirac operator is not given
simply by some gauging of the $\nabla_{\mu}$'s and $\Delta_{\nu}$'s in (\ref{2.2}) but has a more
complicated structure \cite{Kluberg}.}

In light of (\ref{2.3}), when attempting to find a viable $D$ in the free field case 
it is natural to consider Dirac operators of the form
\be
D=\gamma^{\mu}\frac{1}{2a}\nabla_{\mu}+\frac{1}{2a}W +m
\label{2.6}
\ee
on the blocked lattice, where the purpose of $\frac{1}{2a}W$ is to take account of the 
taste-mixing terms in the staggered Dirac operator. In particular, $W$ should formally vanish
$\sim{}a^2$ for $a\to0$, and should lift the species doubling of the naive Dirac operator.
In other words, $\frac{1}{2a}W$ is to be a Wilson-type term.

A feature of the free field staggered Dirac operator (\ref{2.2}) in the taste field representation,
which is very useful in this context, is that $(D_{stt}^{free})^{\dagger}D_{stt}^{free}$
is trivial in spinor$\otimes$flavour space:
\be
(D_{stt}^{free})^{\dagger}D_{stt}^{free}=
\frac{1}{(2a)^2}\left(\,-\nabla^2+\sum_{\nu}({\textstyle \frac{1}{2}}\Delta_{\nu})^2+(2am)^2\right)
({\bf 1}\otimes{\bf 1})
\label{2.7}
\ee
On the other hand, for a free field operator of the form (\ref{2.6}) we have
\be
D^{\dagger}D=
\frac{1}{(2a)^2}\left(\,-\nabla^2+(W+2am)^2\right)
{\bf 1}
\label{2.8}
\ee
trivial in spinor space. Comparing (\ref{2.7}) and (\ref{2.8}), and noting that
$detD=det(D^{\dagger}D)^{1/2}$ (assuming $\frac{1}{2a}W +m\ge0$) and
$det(D_{st})=det(D_{st}^{\dagger}D_{st})^{1/2}$ (assuming $m\ge0$), we immediately
see that a sufficient criteria for the desired determinant relation
$detD=(detD_{st}^{free})^{1/4}$ to be satisfied is
\be
(W+2am)^2=\sum_{\nu}({\textstyle \frac{1}{2}}\Delta_{\nu})^2+(2am)^2\,.
\label{2.9}
\ee
This has the solution
\be
W=\sqrt{(2am)^2+\sum_{\nu}({\textstyle \frac{1}{2}}\Delta_{\nu})^2}\ \ \ -\ 2am\,,
\label{2.10}
\ee
which clearly has the required properties for $\frac{1}{2a}W$ to be Wilson-type term 
(i.e. $W$ lifts species doubling and formally vanishes $\sim{}a^2$ for $a\to0$).
Thus we have arrived at a free field solution $D$ to the SDR (\ref{1.1}).
Substituting (\ref{2.10}) into (\ref{2.6}) we get the expression
\be
D=\gamma^{\mu}\frac{1}{2a}\nabla_{\mu}+\frac{1}{2a}
\sqrt{(2am)^2+\sum_{\nu}({\textstyle \frac{1}{2}}\Delta_{\nu})^2}\,.
\label{2.11}
\ee
Note that turning off the taste-mixing terms in the free field staggered fermion action,
which, as pointed out previously, corresponds to putting $\Gamma^{\nu}\to0$ in (\ref{2.2}),
has the same effect as putting $\Delta_{\nu}\to0$. By (\ref{2.11}) this gives 
$D\to\Sn+m\,$ (for $m\ge0$); thus we have a realisation of the scenario (\ref{2.3}). 
However, because of
the square root in (\ref{2.11}), it is not immediately clear that good locality behaviour of $D$,
anticipated in our earlier discussion, is realised. In fact this square root operator
has some similarity with the free field square root operator considered by Jansen and collaborators 
in \cite{Jansen(NPB)}, which turned out to have unacceptable locality behaviour. 
Nevertheless, we show in the next section that our operator does have good locality behaviour.
The reason why its behaviour is different from the operator in \cite{Jansen(NPB)} is a bit
subtle, and to elucidate this we also provide in the next section a new derivation of the
negative locality result of \cite{Jansen(NPB)} which reveals the origin of the different behaviours.

We remark that other, ultra-local solutions to the SDR exist in the free field case.
Using $-\nabla_{\nu}^2+(\frac{1}{2}\Delta_{\nu})^2=\Delta_{\nu}$ (\ref{2.7}) reduces to
$(D_{stt}^{free})^{\dagger}D_{stt}^{free}=(\frac{1}{(2a)^2}\Delta+m^2)({\bf 1}\otimes{\bf 1})$
and it follows that the free field SDR is satisfied, e.g., by $D=(\frac{1}{(2a)^2}\Delta+m^2)^2$ 
acting on scalar Grassmann fields on the lattice. Other examples of ultra-local solutions are
easily constructed. However, these are unattractive options since they do not have the form of
a lattice Dirac operator.

\section{Free field locality result}

\begin{figure}
$$
\epsfysize=6cm \epsfbox{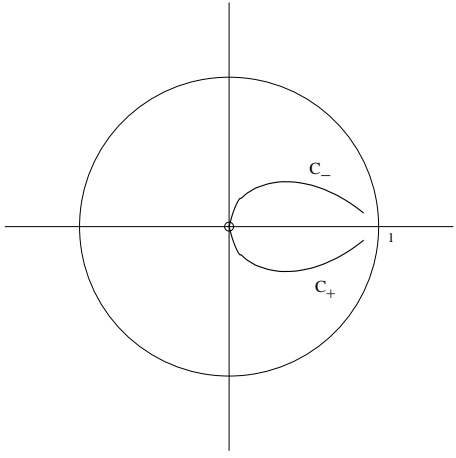}
$$
\caption{The ``exclusion curves'' $C_{\pm}$. The locations of the endpoints
near 1 depend on $am$ and converge to $1$ for $a\to0$.}\label{frp1}\end{figure}

In this section we work in arbitrary spacetime dimension $d$ and show that the free field
operator 
\be
\sqrt{(am)^2+\sum_{\nu}\Delta_{\nu}^2}
\label{3.1}
\ee
on lattice with spacing $a$
is exponentially local with localisation range vanishing $\sim\sqrt{a/m}$ 
for $a\to0$; then $D$ in (\ref{2.11}) obviously has this same locality behaviour on the 
blocked lattice.
The argument proceeds in several steps. First, we specialise to $d\!=\!1$ dimension and write
\be
\sqrt{(am)^2+\Delta^2}\ (x,y)&=&\frac{1}{a}\int_{-\pi}^{\pi}\frac{dp}{2\pi}\,
\sqrt{(am)^2+\Delta(p)^2}\ e^{ip(x-y)/a} \nonumber \\
&=&\frac{1}{2\pi{}ia}\oint_{|z|=1}\frac{dz}{z}\,\sqrt{(am)^2+(2-(z+z^{-1}))^2}\ z^{|x-y|/a}
\nonumber \\
&=&\frac{1}{2\pi{}ia}\oint_{|z|=1}\frac{dz}{z}\,
\sqrt{z^{-2}\Big((am)^2z^2+(z-1)^4\Big)}\ z^{|x-y|/a}
\label{3.2}
\ee
(the integral is counter-clockwise around the unit circle in the complex plane and we have
set $z=e^{\pm{}ip}$ with ``$+$'' if $x-y>0$ and ``$-$'' if $x-y<0\,$; for $x=y$ either
choice can be used.)
The square root $z\mapsto\sqrt{z}$ is holomorphic after making a 
cut in ${\bf C}\,$; we choose the cut to be the half-line of negative real numbers ${\bf R}_-\,$.
Then, by the residue theorem, the circle around which the integral in (\ref{3.2}) is performed
can be shrunk to a closed loop around the region containing the $z$'s for which 
$f(z)\in{\bf R}_-\,$, where
\be
f(z)=z^{-2}\Big((am)^2z^2+(z-1)^4\Big)
\label{3.3}
\ee
is the function inside the square root in (\ref{3.2}), since outside this region (and away from
$z\!=\!0$) $z\mapsto\sqrt{f(z)}$ is holomorphic. 
The excluded $z$'s are are found as follows:
\be
&&\quad f(z)=-\lambda\qquad\ \lambda\in{\bf R}_+ \nonumber \\
&\Leftrightarrow&\ (z-1)^4+((am)^2+\lambda)z^2=0 \nonumber \\
&\Leftrightarrow&\ \Big((z-1)^2+i\sqrt{(am)^2+\lambda}\ \,z\Big)
\Big((z-1)^2-i\sqrt{(am)^2+\lambda}\ \,z\Big)=0
\label{3.4}
\ee
For given $\lambda\in{\bf R}_+$ there are 4 solutions; we are only interested in the ones with
$|z|\le1$ and these are $z=z_{\pm}\Big(\,s\!=\!\sqrt{(am)^2+\lambda}\;\Big)$ where
\be
z_{\pm}=1\pm\frac{is}{2}-\sqrt{s}\sqrt{\pm{}i-s/4}\qquad\quad s\,\in\,\rb\,0,\infty\lb
\label{3.5}
\ee
Thus the $z$'s for which $f(z)\in{\bf R}_-$ and $|z|\le1$ form curves inside the unit circle
in ${\bf C}$, parameterised by (\ref{3.5}) with $s\in\lb\,am,\infty\lb\,$.
It is useful to re-parameterise these curves as follows. We introduce
\be
t=1-\sqrt{\frac{s/2}{s/4+\sqrt{1+(s/4)^2}}}\ \,;
\label{3.6}
\ee
note that this is a strictly decreasing function of $s$ with $t\!=\!1$ for $s\!=\!0$ and
$t\to0$ for $s\to\infty\,$. After a little calculation (\ref{3.5}) can be re-expressed in
terms of $t$ as
\be
z_{\pm}(t)=t\mp i(1-t)\sqrt{\frac{t}{2-t}}\qquad\qquad t\,\in\,\rb\,0,1\lb
\label{3.7}
\ee
The $z$'s for which $f(z)\in{\bf R}_-$ and $|z|\le1$ are now parameterised by the curves
$z_+(t)$ and $z_-(t)$ for $t\in\rb\,0,t_{am}\rb$ where $t_{am}$ is given by setting 
$s=am$ in (\ref{3.6}); we write this out explicitly for future reference:
\be
t_{am}=1-\sqrt{\frac{am/2}{am/4+\sqrt{1+(am/4)^2}}}\ \,.
\label{3.8}
\ee
These curves, which we denote by $C_+$ and $C_-\,$, lie in the lower- and upper half-planes
of ${\bf C}$, respectively. They have a common limit point at $z_+(0)=z_-(0)=0$.
See Fig.\ref{frp1}. 
\indent According to the residue theorem, the integral (\ref{3.2}) remains unchanged when the unit
circle is shrunk to a closed curve $C$ around $C_+\cup\{0\}\cup{}C_-\,$. In the limit this reduces 
to an integral over $C_+\cup\{0\}\cup{}C_-$ itself, with a factor of 2 to take account of the fact 
that $C$ goes along $C_+\cup{}C_-$ twice, with opposite orientations.
(This is assuming that the argument in $z^{|x-y|/a}$ is sufficiently large to
avoid a divergence of the limit integral due to singularity at $z=0\,$; an explicit criterion for 
this will be given further below.)\footnote{If the argument in $z^{|x-y|/a}$ is not sufficiently
large, e.g. if $x=y$, then the curve around which the integral is performed cannot be completely
shrunk to $C_+\cup\{0\}\cup{}C_-$ -- a small detour around $z=0$ must be included. This case
is more subtle, and we do not consider it here since it is not needed to derive the advertised
locality result.} Then, using obvious symmetries,
the integral can be seen to be $2\ \times$ the {\em real part} of the integral over $C_-$.
(The integrals over $C_+$ and $C_-$ are complex conjugate, so the imaginary parts cancel out
as they should.)
The square root $\sqrt{f(z)}$ in the integral then reduces to $\pm{}i\sqrt{\lambda(t)}$
with the explicit expression for $\lambda(t)$ determined below and the sign determined to 
be ``$-$''. Thus (\ref{3.2}) reduces to
\be
\sqrt{(am)^2+\Delta^2}\ (x,y)=\frac{-2}{\pi a}\int_0^{t_{am}}dt\;
\Big|\frac{dz_-}{dt}\Big|\;\sqrt{\lambda(t)}\ \,Re\Big(z_-(t)^{|x-y|/a-1}\Big)
\label{3.9}
\ee
Recalling that the solution to $f(z)=-\lambda$ can be written as (\ref{3.5}) with
$s=\sqrt{(am)^2+\lambda}\;$, and noting that the relation between $s$ and $t$ in (\ref{3.6})
can be inverted to give $s=2(1-t)^2\Big/\sqrt{t(2-t)}\;$, we find
\be
\sqrt{\lambda(t)}=\sqrt{\frac{4(1-t)^4}{t(2-t)}\ -(am)^2}\ \,.
\label{3.10}
\ee
The sign in $\pm{}i\sqrt{\lambda(t)}$ can be determined by considering 
$\sqrt{f(z)}\approx\sqrt{z^{-2}}$ for $z$ near zero. Writing $z=\epsilon+i\delta$ we have
$z^{-2}\approx(-\delta^2-i2\delta\epsilon)/(\delta^4+(2\delta\epsilon)^2)\,$; 
hence, recalling that we have chosen the cut ${\bf R}_-$ to define the square root,
$z^{-2}\stackrel{\epsilon\to0_+}{\to}-i/\delta$. From this it is straightforward to
see that the sign in $\pm{}i\sqrt{\lambda(t)}$ in the integral over $C_-$ is ``$-$'', and this is
the origin of the minus sign in (\ref{3.9}).
Explicit expressions for the remaining ingredients in the integrand in (\ref{3.9}) are readily
found from (\ref{3.7}):
\be
|z_{\pm}(t)|&=&\sqrt{\frac{t}{2-t}} \label{3.11} \\
\Big|\frac{dz_{\pm}}{dt}\Big|&=&\frac{1}{2-t}\,\sqrt{\frac{1+t(2-t)}{t(2-t)}}\ .
\label{3.12}
\ee
Note that the divergences $\sim1/\sqrt{t}$ for $t\to0$ in $\sqrt{\lambda(t)}$ and
$\Big|\frac{dz_-}{dt}\Big|$ are compensated in (\ref{3.9}) by powers of $\sqrt{t}$ in 
$z_-(t)^{|x-y|/a-1}$ provided $|x-y|>a$, which we henceforth assume to be the case.
(This is the criterion alluded to above.)
We can now use (\ref{3.9}) to draw conclusions about the exponential decay of  
$\sqrt{(am)^2+\Delta^2}\ (x,y)$. Explicit evaluation of the integral in (\ref{3.9}) will not
be needed for this, so we do not attempt to perform it here.

For fixed $|x-y|>0$ and given $t\in[0,t_{am}]$ the integrand in (\ref{3.9}) is dominated 
in the $a\to0$ limit by the exponential factor $z_-(t)^{|x-y|/a}$. From (\ref{3.11}) we see that
$|z_-(t)|$ increases with $t$ for $t\in[0,t_{am}]$ (recall $t_{am}\le1$); therefore there can be no
cancellation between the exponential factors for different $t$ in the integral (\ref{3.9})
and it follows that $\sqrt{(am)^2+\Delta^2}\ (x,y)$ decays exponentially 
$\sim z_-(t_{am})^{|x-y|/a}$ for small $a$. From (\ref{3.7}) we see that
\be
t_{am}=1-\sqrt{am/2}+O(am)\,.
\label{3.13}
\ee
Consequently, using (\ref{3.11}), the magnitude of the exponential decay of
$\sqrt{(am)^2+\Delta^2}\ (x,y)$ for small $a$ (i.e. $am<<1$) is found to be
\be
|z_-(t_{am})|^{|x-y|/a}&=&\Big(\,1-\sqrt{am/2}+O(am)\Big)^{|x-y|/2a} \nonumber \\
&=&\left\lb\Big(1-\sqrt{m/2}\,\sqrt{a}+O(am)\Big)^{1/\sqrt{a}}
\;\right\rb^{|x-y|/2\sqrt{a}} \nonumber \\
&\stackrel{a\to0}{\approx}&\Big(e^{-\sqrt{m/2}}\;\Big)^{|x-y|/2\sqrt{a}}
=e^{-\frac{1}{2}\sqrt{\frac{m}{2a}}\ |x-y|}
\label{3.14}
\ee
Thus the localisation range for the exponential decay of 
$\sqrt{(am)^2+\Delta^2}\ (x,y)$ is seen to be $2\sqrt{2a/m}\;$.

We now supplement the preceding with a bound on $|\sqrt{(am)^2+\Delta^2}\ (x,y)|$
which allows to check that the integral in (\ref{3.9}) does not give rise to other factors which
mask the exponential decay when $|x-y|$ is of the same order of magnitude as $\sqrt{a/m}$.
From (\ref{3.10})--(\ref{3.11}) we see that for $t\in[\,0,1]$
\be
\Big|\frac{dz_-}{dt}\Big|\;\le\;\frac{\sqrt{2}}{\sqrt{t}}\quad,\qquad
\sqrt{\lambda(t)}\;\le\;\frac{2}{\sqrt{t}}\quad,\qquad
|z_-(t)|\;\le\;\sqrt{t}
\label{3.15}
\ee
and it follows from (\ref{3.9}) that
\be
\Big|\sqrt{(am)^2+\Delta^2}\ (x,y)\Big|
&\le&\frac{4\sqrt{2}}{\pi{}a}\int_0^{t_{am}}dt\,\Big(\sqrt{t}\;\Big)^{\frac{|x-y|}{a}-3} 
\nonumber \\
&=&\frac{8\sqrt{2}}{\pi\sqrt{t_{am}}\,\Big(|x-y|-a\Big)}
\,\Big(t_{am}\Big)^{|x-y|/2a} 
\label{3.16}
\ee
For $am<<1$ the exponential factor here reduces as in (\ref{3.14}) to give the same decay
found earlier.
The factor $1/\sqrt{t_{am}}$ has no effect since by (\ref{3.13}) it is $\approx1$.
On the other hand, the factor $1/(|x-y|-a)$ blows up for $|x-y|\approx{}a$; however it does not
mask the exponential decay once $|x-y|\ge2a$ (and enhances the locality when 
$|x-y|$ is large).\footnote{Recall that the derivation of (\ref{3.9}), and hence also (\ref{3.16}), 
assumes $|x-y|>a$.} 
When $a$ is sufficiently small, the localisation range ($\sim\sqrt{a/m}\ $) of the exponential
decay is much larger than $a$ and therefore does not get masked by this factor.

We now proceed to the case of arbitrary spacetime dimension $d$ and consider
\be
\sqrt{(am)^2+\sum_{\nu}\Delta_{\nu}^2}\ (x,y)&=&
\frac{1}{(2\pi{}a)^d}\int_{\lb\,-\pi,\pi\rb^d}d^dp\,
\sqrt{(am)^2+\sum_{\nu}\Delta_{\nu}(p_{\nu})^2}\ \,e^{i\sum_{\mu}p_{\mu}(x_{\mu}-y_{\mu})/a}
\nonumber \\
&&\label{3.18}
\ee
Writing $x=(x_1,\bx)\,$, $p=(p_1,\bp)$ and setting
\be
M(\bp)=\sqrt{(am)^2+\sum_{\nu=2}^d\Delta_{\nu}(p_{\nu})^2}
\label{3.19}
\ee
we have
\be
&&\sqrt{(am)^2+\sum_{\nu}\Delta_{\nu}^2}\ (x,y) \nonumber \\
&&\ =\frac{1}{(2\pi{}a)^{d-1}}\int_{\lb\,-\pi,\pi\rb^{d-1}}d^{d-1}\bp\;e^{i\bp\cdot\bx/a}\,
\int_{-\pi}^{\pi}\frac{dp_1}{2\pi a}\,\sqrt{M(\bp)^2+\Delta_1(p_1)^2}\ \,e^{ip_1(x_1-y_1)/a}
\nonumber \\
&&\label{3.20}
\ee
The integral over $p_1$ here is the same as the previous $d=1$ integral (\ref{3.2}) except
that $m$ is replaced here by $M(\bp)$. It can therefore be rewritten as (\ref{3.9}) with
this replacement. By our previous argument this integral decays exponentially
$\sim z_-(t_{M(\bp)})^{|x_1-y_1|/a}$. The decay is slowest when $t_{M(\bp)}$ is largest,
i.e. when $M(\bp)$ is smallest, and this happens when $\bp=(0,\dots,0)$ in which case
$M=M(0)=am$. The same reasoning which led to (\ref{3.14}) then implies that for $am<<1$
the operator kernel $\sqrt{(am)^2+\sum_{\nu}\Delta_{\nu}^2}\ (x,y)$ decays 
$\sim e^{-\frac{1}{2}\sqrt{\frac{m}{2a}}\ |x_1-y_1|}$ along the $\mu\!=\!1$ axis.
Obvious modifications in the preceding show that the same decay holds along any other 
coordinate axis. Thus we see that the localisation range is {\em no smaller than}
$2\sqrt{2}\sqrt{a/m}$. It could however be {\em larger} along directions which are not 
parallel to a coordinate axis. To derive an upper bound on the localisation range we use bounds
similar to those leading to (\ref{3.16}) to get\footnote{The factor $\frac{1}{a^{d-1}}$ originates
from the first integral in (\ref{3.20}):
$\frac{1}{(2\pi a)^{d-1}}\int_{\lb\,-\pi,\pi\rb^{d-1}}d^{d-1}\bp\;|e^{i\bp\cdot\bx/a}|
=\frac{1}{a^{d-1}}$.}
\be
\Big|\sqrt{(am)^2+\sum_{\nu}\Delta_{\nu}^2}\ (x,y)\Big|\;\le\;
\frac{8\sqrt{2}}{a^{d-1}\pi\sqrt{t_{am}}
\;(|x_{\mu}-y_{\mu}|-a)}
\;\Big(t_{am}\Big)^{|x_{\mu}-y_{\mu}|/2a} 
\label{3.21}
\ee
holding for each $\mu=1,2,\dots,d$. It follows that 
\be
\Big|\sqrt{(am)^2+\sum_{\nu}\Delta_{\nu}^2}\ (x,y)\Big|^d\;\le\;\prod_{\mu=1}^d\;
\frac{8\sqrt{2}}{a^{d-1}\pi\sqrt{t_{am}}
\;(|x_{\mu}-y_{\mu}|-a)}
\;\Big(t_{am}\Big)^{|x_{\mu}-y_{\mu}|/2a} 
\nonumber
\ee
which in turn gives
\be
\Big|\sqrt{(am)^2+\sum_{\nu}\Delta_{\nu}^2}\ (x,y)\Big|&\le&
\frac{8\sqrt{2}}{a^{d-1}\pi\sqrt{t_{am}}
\;\Big(\prod_{\mu}(|x_{\mu}-y_{\mu}|-a)\Big)^{1/d}}
\;\Big(t_{am}\Big)^{||x-y||_1/2da} 
\nonumber \\
&&\label{3.22}
\ee
where $||x-y||_1=\sum_{\mu}|x_{\mu}-y_{\mu}|$ is the `taxi-driver' norm.
A calculation analogous to (\ref{3.14}) gives
\be
\Big(t_{am}\Big)^{||x-y||_1/2da}\ \stackrel{a\to0}{\approx}\
e^{-\frac{1}{2d}\sqrt{\frac{m}{2a}}\ ||x-y||_1}\,.
\label{3.23}
\ee
Since $||x-y||\le||x-y||_1$ it follows from this and (\ref{3.22}) that the localisation range is
{\em no bigger} than $2d\sqrt{2a/m}\,$, i.e. it lies between this value and the previously
derived lower limit $2\sqrt{2a/m}$. This completes the demonstration of exponential locality,
with localisation range vanishing $\sim\sqrt{a/m}\,$, claimed at the beginning of this section.

It is interesting to compare this result with the free field locality result derived in 
\cite{Jansen(NPB)} for the operator
\be
\sqrt{(am)^2+\sum_{\nu}(\nabla_{\nu})^{\dagger}\nabla_{\nu}}\,.
\label{3.24a}
\ee
This operator was shown there to be exponentially local but with localisation range remaining
finite in the $a\to0$ limit. The argument involved a mixture of analytic and numerical 
calculations\footnote{Specifically, the locality behaviour of the {\em continuum version} of this
operator was analytically determined and numerical calculations were then performed to 
check that the lattice operator kernel reduced to the continuum expression in the $a\to0$
limit -- see Part 3 of Appendix B in \cite{Jansen(NPB)}.}; however, the result can be
established by purely analytic means, using the techniques introduced in the preceding, as
we now demonstrate. This will also show the origin of the difference in locality behaviour
between our operator and this one. 
(Note $\sum_{\nu}\nabla_{\nu}^{\dagger}\nabla_{\nu}=-\nabla^2\,$; we use the latter expression
in the following.)

\begin{figure}
$$
\epsfysize=6cm \epsfbox{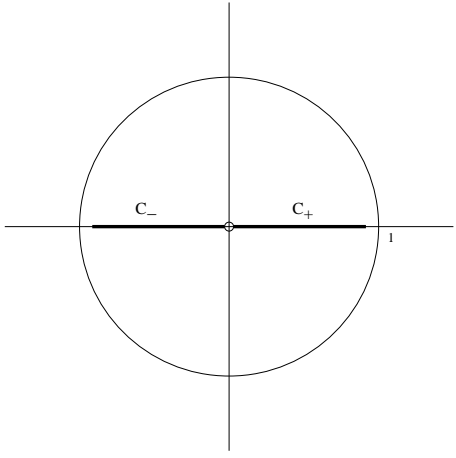}
$$
\caption{The ``exclusion curves'' $C_{\pm}$ for the operator (\ref{3.24}). Compare with
Fig. \ref{frp1} for the operator (\ref{3.2}).}\label{frp2}\end{figure}

In the $d\!=\!1$ case,
\be
\sqrt{(am)^2-\nabla^2}\ (x,y)&=&\frac{1}{2\pi a}\int_{-\pi}^{\pi}dp\,\sqrt{(am)^2-\nabla^2(p)}
\ \,e^{ip(x-y)/a} \nonumber \\
&=&\frac{1}{2\pi ia}\oint_{|z|=1}\frac{dz}{z}\,\sqrt{(am)^2-(z-z^{-1})^2}\ \,z^{|x-y|/a}
\label{3.24}
\ee
Setting $g(z)=(am)^2-(z-z^{-1})^2=z^{-2}\Big((am)^2z^2-(z^2-1)^2\Big)$ we proceed as before
by determining the $z$'s satisfying $g(z)\in{\bf R}_-$ and $|z|\le1\,$:
\be
&&\quad g(z)=-\lambda\qquad\ \lambda\in{\bf R}_+ \nonumber \\
&\Leftrightarrow&\ -(z^2-1)^2+((am)^2+\lambda)z^2=0 \nonumber \\
&\Leftrightarrow&\ \Big(z^2-1+\sqrt{(am)^2+\lambda}\ \,z\Big)
\Big(z^2-1-\sqrt{(am)^2+\lambda}\ \,z\Big)=0
\label{3.25}
\ee
The solutions with $|z|\le1$ are $z_{\pm}\Big(s=\sqrt{(am)^2+\lambda}\;\Big)$ where
\be
z_+(s)=-s/2+\sqrt{1+(s/2)^2}=-z_-(s)\qquad\quad s\,\in\,\lb\,0,\infty\lb
\label{3.26}
\ee
Note that $z_{\pm}(s)\to0$ for $s\to\infty$. Hence the solutions form curves $C_+$ and $C_-$
inside the unit circle in ${\bf C}$, parameterised, respectively, by $z_+(s)$ and $z_-(s)\,$,
$\;s\in\lb\,am,\infty\lb$. The curves in this case are simply intervals on the real axis:
$C_+=\rb\,0,z_+(am)\rb$ and $C_-=\lb\,-z_+(am),0\lb\,$ (see Fig.\ref{frp2}).
The circle around which the integration in (\ref{3.24}) is carried out can now be shrunk to a 
closed curve around $C_+\cup\{0\}\cup C_-\,$, leading in the limit to an integral over these curves.
By arguments similar to those in the previous case one then finds that 
$\sqrt{(am)^2-\nabla^2}\ (x,y)$ decays exponentially $\sim|z_+(am)|^{|x-y|/a}$.
From (\ref{3.26}) we see that for $am<<1$ the magnitude of the decay factor becomes
\be
|z_+(am)|^{|x-y|/a}&=&\Big((1-am/2+O((am)^2))^{1/a}\Big)^{|x-y|} \nonumber \\
&\stackrel{a\to0}{\approx}&e^{-\frac{m}{2}\,|x-y|}\,.
\label{3.27}
\ee
Thus the localisation range in this case, $2/m$, is independent of $a$ and remains finite in the
$a\to0$ limit. The general dimension $d$ case can now be dealt with by an argument analogous to
our earlier one; this leads to the result that $\sqrt{(am)^2-\nabla^2}\ (x,y)$ is exponentially
local with lower- and upper bounds on the localisation range being $2/m$ and $2d/m$, respectively,
showing that the range is also finite 
$\sim m^{-1}$ in the $a\to0$ limit in the general dimension $d$ case.
Thus we reproduce the general finding of \cite{Jansen(NPB)} for the locality behaviour of this free
field operator.\footnote{In the expression for the free field operator kernel $G(x,y)$ in 
Eqn.(3.11) of \cite{Jansen(NPB)} the integration range for $dp_{\mu}$ (after a change of
variables $p_{\mu}\to p_{\mu}/a$) is $\lb\,-\pi/2,\pi/2\rb$. But since 
$-\nabla_{\mu}^2(p_{\mu})=sin^2(p_{\mu})$ this gives precisely 1/2 of what the integration
over $\lb\,-\pi,\pi\rb$ would give. Hence $G(x,y)=\frac{1}{2^d}\sqrt{(am)^2-\nabla^2}\ (x,y)$
so the locality result derived above applies.}

The origin of the different locality behaviour of our operator and the free field operator 
considered in \cite{Jansen(NPB)} is now apparent: The exponential decay in $d\!=\!1$ dimension,
which, as we have seen, is the same as the decay along a coordinate axis in general $d$ 
dimensions, is given in both cases by $|z_{max}|^{|x-y|/a}\,$, where $z_{max}$ is the point 
on the ``exclusion curves'' in Fig.\ref{frp1} (our case) 
or Fig.\ref{frp2} (the case of Ref.\cite{Jansen(NPB)})
which is closest to the unit circle. In our case, 
\be
|z_{max}|\;\approx\;(t_{am})^{1/2}\;\approx\;\Big(1-\sqrt{am/2}+O(am)\Big)^{1/2}\,,
\label{3.28}
\ee 
resulting in decay $\sim e^{-\frac{1}{2}\sqrt{\frac{m}{2a}}\ |x-y|}\,$, whereas in the case of 
Ref.\cite{Jansen(NPB)}, 
\be
|z_{max}|\;\approx\;1-am/2+O((am)^2)\,,
\label{3.29}
\ee
resulting in decay $\sim e^{-\frac{m}{2}\,|x-y|}$. The essential difference is that the leading 
$a$-dependent term inside the bracket in (\ref{3.28}) is $\sim\sqrt{am}\,$ whereas the 
corresponding leading term in (\ref{3.29}) is $\sim am$.
This leads to the localisation range being $\sim\sqrt{a/m}$ in the former case and $\sim m$
(independent of $a$) in the latter. The former tends to zero for $a\to0$ while the latter
stays constant in this limit.

We remark that the technique of writing the kernel $\O(x,y)$ of a 
free field lattice operator as an
integral around a closed curve in the complex plane, and then attempting to shrink the curve
as a way of deriving locality properties, was used previously in a different context
in Ref.\cite{Fujikawa}.

\section{Discussion}

Working in the flavour (taste) representation for staggered fermions, 
we have shown that in the free field case there is a simple and natural Wilson-type
lattice Dirac operator $D$ on the blocked lattice, given by (\ref{2.11}), which satisfies 
$detD=(detD_{staggered})^{1/4}$ and is exponentially local with localisation range vanishing
$\sim\sqrt{a/m}\;$ for $a\to0$. The techniques developed to derive the free field locality 
result are of a generally applicable nature, and we also used them to give a new, purely
analytic derivation of the negative locality result in \cite{Jansen(NPB)}. They can also be
used to study free field locality properties of other lattice operators of current interest;
in particular, the overlap Dirac operator \cite{Neu1}, which is treated in a forthcoming 
paper \cite{Bicudo}.

Our free field operator can be gauged, i.e. coupled to the link variables of the original
lattice, in a variety of ways. The simplest way is to define link variables $V_{\mu}$ 
on the blocked lattice in terms of the link variables $U_{\mu}$ on the original lattice by
\be
V_{\mu}(2x)=U_{\mu}(2x)U_{\mu}(2x+a\hat{\mu})
\label{4.1}
\ee
($\hat{\mu}$=unit vector in the positive $\mu$-direction); then the difference operators
$\nabla_{\mu}\,$, $\Delta_{\nu}$ on the blocked lattice in (\ref{2.11}) can be coupled to 
$V_{\mu}$ in the usual way -- this specifies the ``minimal gauging'' of our $D$. 
However, the resulting operator cannot be expected to satisfy the SDR in the interacting case.
Our argument in \S2 does not carry over to this case; it is specific to the free field case.

Regarding the possibility of gauging our operator such that the SDR does continue to hold in the
interacting case, we note the following. The taste-mixing part of the staggered Dirac operator
leaves unbroken a U(1) subgroup of the continuum U(4) axial flavour symmetry.\footnote{We are
assuming that the mass term of the staggered fermion is of the standard form 
$m({\bf 1}\otimes{\bf 1})$. (If the mass matrix is not proportional to the identity operator then
the interpretation of the U(1) symmetry is different.) Note that the U(1) symmetry is {\em not}
the diagonal U(1) subgroup in U(4). The latter, associated with the axial anomaly, is explicitly
broken by the taste-mixing part of $D_{staggered}$.} 
In the taste representation, this symmetry can be expressed in the free field case as
\be
\lbrace\,\gamma_5\otimes\Gamma_5\,,\,D_{stt}\rbrace=0\qquad\quad\mbox{($m=0$)}
\label{4.2}
\ee
with notations as in (\ref{2.2}). This chiral symmetry protects staggered fermions against
additive mass renormalisation \cite{Weiss,Mitra,Golterman}. The sea quark effective action in
dynamical staggered fermion simulations is $log\,det(D_{st})^{1/4}=\frac{1}{4}Tr\,logD_{st}\,$,
the same as for a usual staggered fermion modulo an overall factor $1/4$. Thus the protection 
against additive mass renormalisation is also present in this case. This situation would be 
difficult, if not impossible, to reconcile with the existence of a single-flavour $D$ satisfying 
the SDR unless the $D$ is also protected against additive mass renormalisation: if the bare mass
is small then for staggered fermions the physical mass will also be small, whereas for a lattice
fermion described by a $D$ which does not have a chiral symmetry the physical mass will be large 
due to the additive mass renormalisation induced by radiative corrections. Since our $D$ is of 
Wilson-Dirac form, any gauged version of it will be afflicted with additive mass renormalisation,
unless there is a very special choice of gauging which endows this $D$ with a new, hitherto
undiscovered type of chiral symmetry. The latter seems very unlikely though, so most probably a
gauged version of our $D$ satisfying the SDR simply does not exist.

The symmetry corresponding to (\ref{4.2}) for $(D_{stt})^{\dagger}D_{stt}$ in the free field 
case is
\be
\lb\,\gamma_5\otimes\Gamma_5\,,\,(D_{stt})^{\dagger}D_{stt}\rb=0\qquad\quad\mbox{($m=0$)}\,.
\label{4.3}
\ee
But, as noted in \S2, $(D_{stt})^{\dagger}D_{stt}\sim{\bf 1}\otimes{\bf 1}$ in the free field 
case, so (\ref{4.3}) is trivially satisfied. This explains why it was possible to find a 
single-flavour $D$ without chiral symmetry but nevertheless satisfying the SDR in the free field 
case. In the interacting case $(D_{stt})^{\dagger}D_{stt}$ is no longer 
$\sim{\bf 1}\otimes{\bf 1}$ and the gauged version of the symmetry (\ref{4.3}) is a nontrivial
property. 

Although a gauged version of our $D$ satisfying the SDR is unlikely to exist, the 
free field locality result
for it is still relevant as a general indication of the possibility of having
local single-flavour Dirac operators satisfying the SDR, and as a first step toward constructing
more sophisticated operators which have a chance to be local (also for $m\!=\!0$) and satisfy
the SDR in the full interacting case. At present it is the only analytic
positive locality result derived for
a solution of the SDR in any gauge background (the background in our case being the trivial one).

The above discussion indicates that, for a viable $D$ to satisfy the SDR
in the full interacting case, it should have an exact chiral-type symmetry (presumably 
corresponding in some way to the aforementioned chiral symmetry of staggered fermions).
The only such symmetry currently known for single-flavour lattice Dirac operators is the
lattice-deformed chiral symmetry \cite{Luscher(PLB)} possessed by operators satisfying the 
Ginsparg-Wilson (GW) relation \cite{GW,Laliena,Neu2} and its generalisations \cite{Fuji(GW)}.
This suggests to look for free field solutions to the SDR which also satisfy the GW relation,
in the hope that among these there may be a $D$ which can be gauged such that the SDR continues 
to hold in the interacting case. In fact this was already investigated by numerical means on 
finite lattices in Ref.\cite{Peardon}. The numerical results there appear to be encouraging.
However, the problem can also be addressed analytically and when this is done difficulties 
become apparent. Setting the lattice spacing of the blocked lattice to unity for convenience, 
the GW relation in its broad form is 
\be
\gamma_5D+D\gamma_5=2D\gamma_5RD
\label{4.4}
\ee
where $R$ is an arbitrary local scalar hermitian operator. As noted at the end of \S2, 
a sufficient condition for the free field $D$ to satisfy the SDR is
\be
D^{\dagger}D=\Delta+m^2
\label{4.5}
\ee
In the free field case, assuming $\gamma_5$-hermiticity $D^{\dagger}=\gamma_5D\gamma_5$
and exploiting the fact that solutions to (\ref{4.4}) are of the form
$D=(2R)^{-1}(1+\gamma_5\epsilon)$ where $\epsilon^2=1$ and $\epsilon^{\dagger}=\epsilon\,$, 
the most general solution to (\ref{4.4}) is seen to be of the form
\be
D_{GW}=\frac{1}{2R}\bigg(\,1+\frac{\gamma^{\mu}\wnabla_{\mu}+W}{\sqrt{-\wnabla^2+W^2}}\,\bigg)
\label{4.6}
\ee
where $\wnabla_{\mu}$ and $W$ are sums of scalar operators multiplied by an even number of 
$\gamma$-matrices. Straightforward algebra now shows that requiring this operator to satisfy
(\ref{4.5}) fixes the $W$ such that
\be
D_{GW}\;=\;\gamma^{\mu}\wnabla_{\mu}\,\frac{R}{|R|}
\sqrt{\frac{(\Delta+m^2)(1-\frac{1}{4}(2R)^2(\Delta+m^2))}
{-\wnabla^2}}\ +\ R(\Delta+m^2)
\label{4.7}
\ee
This is the general solution to (\ref{4.4})--(\ref{4.5}) in the free field case. 
The numerical solutions
investigated in Ref.\cite{Peardon} are particular cases of this operator (or more precisely,
approximations to it on finite lattices) with the $\wnabla_{\mu}$'s being scalar operators.
The main interest here is in the case $m\!=\!0$, since it is in the chiral limit that $D$
should have the chiral symmetry implied by the GW relation (\ref{4.4}). 

In the simplest case where $R=1/2$ and
$\wnabla_{\mu}=\nabla_{\mu}\,$, i.e. the usual symmetrised difference 
operator, $D_{GW}$ reduces in the $m\!=\!0$ case to
\be
\gamma^{\mu}\nabla_{\mu}
\sqrt{1-{\textstyle \frac{1}{4}}\sum_{\mu\ne\nu}\Big(\frac{\Delta_{\mu}\Delta_{\nu}}{-\nabla^2}
\Big)}\ \;+\ {\textstyle \frac{1}{2}}\Delta
\label{4.8}
\ee
While this operator correctly reproduces the continuum free field Dirac operator in the $a\to0$
limit, it is most unlikely to have acceptable locality behaviour. The presence of the 
$(-\wnabla^2)^{-1}$ inside the square root in the general operator (\ref{4.7}) makes it difficult
to envisage that there exist $\wnabla_{\mu}$'s and a local $R$ for which this operator has 
acceptable locality behaviour either, in spite of the numerical indications from Ref.\cite{Peardon}.
Thus it would seem that the condition (\ref{4.5}), which is sufficient, but not necessary, for the 
SDR to be satisfied, is actually too restrictive to lead to a local operator $D$ satisfying both
the SDR and GW relation.

The preceding considerations indicate that finding a viable exact solution to the SDR in the 
interacting case is a difficult problem. However, to resolve the fourth root issue is is not
actually necessary to have an exact solution; it suffices to find a viable lattice Dirac operator
which satisfies the SDR {\em approximately} in the sense that the effective action difference
\be
d(U)={\textstyle \frac{1}{4}}\,log\,detD_{st}-log\,detD
\label{4.11}
\ee
is effectively just a lattice Yang-Mills action for the gauge field. In this case, representing
the quark determinant by $det(D_{st})^{1/4}$ is physically equivalent to representing it
by $detD$ together with a renormalisation of the bare coupling constant (i.e. a shift in 
$\beta$). In other words, $d(U)$ has the same effect as the fermion determinant for dynamical
heavy quarks \cite{Hasenfratz-DeGrand}. In connection with this it is useful to note that the 
perturbative expansion of a general single-flavour lattice fermion determinant has the form
\cite{DA(prep)}
\be
log\,detD&=&(-{\textstyle \frac{1}{8\pi^2}}\,log(am)^2+c_D)\,S_{YM}(A)
+\sum_{n=2}^{\infty}(I_n(A;m)+v_n(A;am)) \nonumber \\
&&\label{4.12}
\ee
where $S_{YM}(A)$ is the continuum Yang-Mills action, $I_n(A;m)$ is a non-local continuum 
functional of order $n$ in $A$, and the $v_n(A;am)$'s (also nonlocal and of order $n$ in $A$)
are terms which vanish for $am\to0$. The dependence on
the specific choice of lattice Dirac operator $D$ enters only through the numerical coefficient
$c_D$ and the functions $v_n(A;am)$. (A gauge field-independent term which diverges
for $am\to0$ has been ignored in (\ref{4.12}).) In fact the perturbative expansion
of $\frac{1}{4}log\,detD_{st}$ has the same form (\ref{4.12}) as a single-flavour Dirac operator
\cite{DA(prep)}. Letting $c_{st}$ denote the coefficient $c_D$ in this case, it follows that
the perturbative expansion of the effective action difference (\ref{4.11}) has the form
\be
d(A)=(c_{st}-c_D)S_{YM}(A)+\sum_{n=2}^{\infty}w_n(A;am)
\label{4.13}
\ee
with each $w_n(A;am)$ vanishing for $am\to0$. Thus it would seem that, in the perturbative setting
at least, for $am<<1$ the effective action difference is indeed just a Yang-Mills action for the 
gauge field, for any sensible choice of single-flavour $D$. The situation is not this simple 
though -- although they vanish for $a\to0$, the functions $v_n$ in (\ref{4.12}) and $w_n$ in
(\ref{4.13}) still do affect the quantum continuum limit. At the perturbative level this
is manifested in that Feynman diagrams with vertices from these terms can be non-vanishing; 
in fact divergent. To see this, recall that
the terms in the perturbative expansion of $log\,detD$ are given by 
1-(fermion)-loop gluonic $n$-point functions. The internal propagators and vertices in these
receive radiative corrections. In particular, unless $D$ is protected by a chiral symmetry, the
radiative corrections to the internal propagators give rise to a large additive mass 
renormalisation. Since there is no corresponding effect from $D_{st}$ to cancel this, it will 
manifest itself in the divergence (for $am\to0$) of various Feynman diagrams involving vertices
from the ``irrelevant'' terms $w_n(A;am)$ in the effective action difference (\ref{4.13}).
Thus the importance of $D$ having a chiral symmetry becomes clear in this context as well.

In perturbative (lattice) QCD the renormalisations of interaction vertices are independent
of the renormalisation of the fermion propagator and bare mass. Thus (\ref{4.13}) and the 
observations above would suggest that, for $am<<1$, the effective action difference $d(A)$ 
is indeed essentially a Yang-Mills action when the single-flavour lattice Dirac operator $D$
has an exact chiral symmetry; e.g., when $D$ is the overlap Dirac operator. Support for this
hypothesis comes from a numerical study carried out in two dimensions in Ref.\cite{Durr}. 
In two spacetime dimensions the perturbative expansion of $log\,detD$ is completely
universal, modulo terms which vanish for $am\to0$ \cite{DA(prep)} (this is a reflection of the
fact that QCD in two dimensions is super-renormalisable). Thus the first term in the right-hand
side of the effective action difference (\ref{4.13}) is absent in this case, and the hypothesis
then states that representing the quark determinant by $det(D_{st})^{1/2}$ is equivalent
to using $detD_{ov}$ (with $D_{ov}$ being the overlap Dirac operator) {\em without} any 
renormalisation of the bare coupling. If this hypothesis holds, then $det(D_{st})^{1/2}$ should
coincide with $detD_{ov}$ for equilibrium gauge configurations of an ensemble generated
by taking the probability weight to be $e^{-\beta S_{YM}(U)}\,det(D_{st})^{1/2}$.
And this is precisely what was found to good accuracy in a numerical study in Ref.\cite{Durr}.

It must be remembered though that the perturbative picture is not the full picture. Low-lying
eigenvalues of the Dirac operator are associated with long-range, low energy dynamics in QCD
which is not captured by the perturbative framework. Indeed, numerical studies 
of the Wilson fermion determinant in Ref.\cite{Duncan} show that the log of the determinant 
cannot be modelled by a linear combination of local loop functionals (i.e. functional of the 
form $Tr\,U(\sigma)$ where $U(\sigma)$ is the product of the link variables around a closed lattice
path $\sigma$); in particular it cannot be modelled by a local lattice YM action. 
However, the product of the Dirac eigenvalues of magnitude $\ge\Lambda_{QCD}$ {\em does} admit 
such a description \cite{Duncan}.
Thus, the aforementioned hypothesis, coming from the perturbative considerations above, should be
regarded as applying to the {\em truncations} of $det(D_{st})^{1/4}\,$, $detD$, and the effective 
action difference $d(U)$, given by excluding the eigenvalues of magnitude $<\Lambda_{QCD}$.
(A way to implement and study this truncation in the perturbative setting is mentioned in 
\cite{Duncan1}.)

Specifically, defining $detD_{high}$ and $detD_{low}$ to be the products of the eigenvalues of $D$ 
of magnitudes $\ge\Lambda_{QCD}$, and $<\Lambda_{QCD}$, respectively, and splitting up the
effective action difference into $d=d_{high}+d_{low}$ in the obvious way, the hypothesis can
be stated as follows:
``When $D$ is the overlap Dirac operator and $am\!<\!<\!1$ then $d_{high}(U)$ is essentially a 
local lattice YM action in 4 dimensions, and essentially vanishing in 2 dimensions.''
The question of whether $d(U)$ itself is effectively a local YM action (which
would give a positive resolution of the fourth root issue if the answer is
affirmative) is then reduced to a question of whether or not $d_{low}(U)$ is effectively zero.
Thus it would be highly desirable to numerically study 
$det(D_{st,low})^{1/4}\,$, $detD_{ov,low}\,$, 
and thereby $d_{low}(U)$, in equilibrium gauge backgrounds in 4 dimensions. We remark that, in 
2 dimensions, combining the hypothesis that $d_{high}$ vanishes (in 2 dim.) with the numerical
agreement \cite{Durr} between the full rooted staggered and overlap determinants implies that 
$d_{low}$ does indeed vanish in this case.

In 4 dimensions numerical studies have found that, after applying a UV-filtering procedure, there 
is good agreement between the low-lying eigenvalues of $D_{ov}$ and $D_{st}$ (modulo a four-fold 
degeneracy in the latter)\cite{Durr}.\footnote{In fact 
it is only when this UV-filtering is applied that
the aforementioned agreement between the rooted staggered and overlap determinants in 2 dim.
holds \cite{Durr}. Without the filtering the agreement breaks down, just as the agreement
between the low-lying eigenvalues does. So it is tempting to ascribe the breakdown in the
agreement between the determinants to the breakdown in the agreement between the low-lying 
eigenvalues. This gives a further hint that vanishing of $d_{low}$ is intimately connected
with having agreement between the low-lying eigenvalues of $D_{ov}$ and $D_{st}$.} 
Comparisons of the spectrum of $D_{st}$ with 
predictions of Random Matrix Theory also back up this picture 
\cite{Follana,Woloshyn}. While this does not
by itself prove that $d_{low}$ vanishes, it is certainly compatible and suggestive of it.

While the numerical work in this direction may lead to a resolution of the fourth root issue
at a practical level, one should ask whether it is possible to also get a resolution at
the theoretical level in this approach. If it is possible it will probably happen as follows:
(i) Use renormalisation group arguments to justify a perturbative treatment of $d_{high}$ and 
verify the hypothesis that it is effectively a local lattice YM action. (ii) By applying 
Random Matrix Theory and theoretical implications of UV-filtering
to the low-lying spectra of $D_{ov}$ and $D_{st}$ show that $d_{low}$ is effectively 
zero when taking the quantum continuum limit.

Another interesting and promising approach to the fourth root issue has been given recently
by Shamir \cite{Shamir}. A renormalisation group argument is used to express the free field 
staggered fermion action in the flavour (taste) representation on a lattice spacing $a_0$ 
as an action on a coarse lattice of spacing $a=2^na_0\,$. This results in a decomposition of the
staggered fermion determinant in the form
\be
det(D_{st})=det(D_n)\,det(G_n^{-1})
\label{4.14}
\ee
The operator $D_n$ encodes the low energy/long range dynamics of staggered fermions;
it decays exponentially with localisation range $\sim a\,$, satisfies a GW relation when 
$m\!=\!0$, and becomes proportional to the identity matrix in flavour space in the large $n$ limit:
$\lim_{n\to\infty}D_n=D_{rg}\otimes{\bf 1}_{flavour}\;$. Hence, in this limit
\be
det(D_{st})^{1/4}=det(D_{rg})\,det(G_{\infty}^{-1/4})\,.
\label{4.15}
\ee
The magnitude of the spectrum of each $G_n^{-1}$ has a lower bound $\sim a$; in units of the fine
lattice spacing $a_0$ this blows up for large $n$, so the expectation is that in a gauged version
of this setting the effect of $det(G_{\infty}^{-1/4})$ in (\ref{4.15}) is exactly the same as 
that of the determinant of a heavy dynamical fermion: to simply renormalise the bare
coupling parameter.

It should be pointed out though that the $n\to\infty$ limit leading to (\ref{4.15})
cannot actually be taken in practise -- it corresponds to $a\to0$, but $a$ must remain
non-zero since it is the spacing of the lattice on which the staggered fermion lives and
the Lattice QCD simulations are performed. Therefore, in this approach one needs to remain
at finite $n$, i.e. the setting of (\ref{4.14}). For large finite $n$ the operator $D_n$ is
close to being diagonal in flavour space, but is not exactly diagonal. This is different from
the situation in the present paper where we obtain a single-flavor candidate Dirac operator
already at non-zero lattice spacing. To fully resolve the fourth root issue in Shamir's
approach it is necessary to find an single-flavour lattice Dirac operator $D'$ such that
adding $log\,detD'-\frac{1}{4}log\,detD_n$ to the lattice gauge field action does not affect
the quantum continuum limit. Shamir has a proposal for this operator $D'$ \cite{Shamir(private)}.
Moreover, his approach has a definite possibility of being extended to the interacting case,
although this remains a difficult challenge for future work.

An appealing feature of Shamir's approach is that the GW chiral symmetries of $D_n$ 
and $D_{rg}$ (at $m\!=\!0$)
originate in a clear and direct way from the chiral symmetry (\ref{4.2}) of the staggered
Dirac operator. This also raises intriguing questions. The chiral symmetry of a GW Dirac operator
is generally anomalous -- it gets broken by the fermion integration measure \cite{Luscher(PLB)}.
On the other hand, the chiral symmetry of $D_{st}$ gets broken spontaneously in the $m\to0$ limit
(at least at strong coupling), and there is an associated Goldstone meson \cite{Wolff,Sch}.
In connection with this we mention a potentially troubling aspect of the fourth root prescription
which has been pointed out already by Creutz \cite{Creutz}: When the determinant for a single 
quark is represented by $det(D_{st})^{1/4}\,$, what becomes of the Goldstone meson
associated with the spontaneous breaking of the chiral symmetry of $D_{st}$? Single-flavour Dirac
operators are not supposed to have spontaneously broken chiral symmetries. 
This and other intriguing issues for the fourth root prescription remain as an urgent topic for 
future work.

Finally, we mention that a completely different approach to this issue, 
involving relating the fourth root prescription to local theories via a parameter 
deformation in a family of lattice theories in 6 spacetime dimensions, has been
described by Neuberger in Ref.\cite{Neu(root)}.

{\bf Acknowledgements}. I thank Mike Peardon for an interesting discussion and for telling me about
Ref.\cite{Peardon} prior to publication. Thanks also go to the National Center for Theoretical
Sciences and nuclear/particle physics group at National Taiwan University for kind hospitality 
and support during visits where part of this work was done, and, in particular, to Ting-Wai Chiu 
for making the arrangements. For interesting discussions on topics related to this work I would
like to thank Mike Creutz, Urs Heller,
Taku Izubuchi, Tony Kennedy, Peter Orland and, in particular, Rajamani 
Narayanan, who I also thank for drawing my attention to Ref.'s \cite{Duncan,Duncan1}.
Last but not least I thank Yigal Shamir for useful correspondence.
At Leiden the author was supported by a Marie Curie fellowship from 
the European Commission, contract HPMF-CT-2002-01716. Some of the work on this manuscript was done 
during the KITP program ``Modern Challenges for Lattice Field Theory'' and I thank the KITP for 
support under NSF Grant No. PHY99-0794. At FIU the author is supported by NSF Grant PHY-0400402.


\begin{thebibliography}{XXX}

\bibitem{improved}
C. W. Bernard {\em et. al.}, Nucl. Phys. Proc. Suppl. 60A, 297 (1998);
G. P. Lepage, Nucl. Phys. Proc. Suppl. 60A, 267 (1998) [hep-lat/9707026];
C. W. Bernard {\em et. al.}, [MILC Collaboration], Phys. Rev. D58, 014503 (1998) [hep-lat/9712010];
G. P. Lepage, Phys. Rev. D59, 074502 (1999) [hep-lat/9809157];
K. Orginos and D. Toussaint [MILC], Nucl. Phys. Proc. Suppl. 73, 909 (1999) [hep-lat/9809148];
K. Orginos, D. Toussaint and R. L. Sugar [MILC], Phys. Rev. D60, 054503 (1999) [hep-lat/9903032];
C. W. Bernard {\em et. al.}, [MILC], Phys. Rev. D61, 111502 (2000) [hep-lat/9912018];
K. Orginos and D. Toussaint [MILC], Phys. Rev. D59, 014501 (1999) [hep-lat/9805009].

\bibitem{Davies(PRL)}
C.T.H. Davies {\em et. al.} [HPQCD, UKQCD, MILC and Fermilab Collaborations],
Phys. Rev. Lett 92, 022001 (2004) [hep-lat/0304004].

\bibitem{Bernard}
C. Aubin {\em et. al.} [MILC], Phys. Rev. D70, 114501 (2004) [hep-lat/0407028].

\bibitem{Allison}
I.F. Allison {\em et. al.} [HPQCD, UKQCD and Fermilab Collaborations],
Phys. Rev. Lett. 94, 172001 (2005)[hep-lat/0411027].

\bibitem{Jansen(L2003)}
K. Jansen, Nucl. Phys. Proc. Suppl. 129, 3 (2004) [hep-lat/0311039].

\bibitem{DeGrand}
T. DeGrand, Int. J. Mod. Phys. A19, 1337 (2004) [hep-ph/0312241].

\bibitem{Neu}
H. Neuberger, hep-ph/0402148.

\bibitem{DA}
D.H. Adams, Phys. Rev. Lett. 92, 162002 (2004) [hep-lat/0312025];
hep-lat/0409013.

\bibitem{Jansen(NPB)}
B. Bunk, M. Della Morte, K. Jansen and F. Knechtli, Nucl. Phys. B697, 343 (2004) [hep-lat/0403022].

\bibitem{Jansen(L2004)}
B. Bunk, M. Della Morte, K. Jansen and F. Knechtli, hep-lat/0408048.

\bibitem{Hart}
A. Hart and E. Muller, hep-lat/0406030.

\bibitem{Neu(root)}
H. Neuberger, Phys. Rev. D70:097504 (2004) [hep-lat/0409144].

\bibitem{Peardon}
F. Maresca and M.J. Peardon, {\em A path integral representation of the free one-flavour
staggered-fermion determinant.}, hep-lat/0411029.

\bibitem{Shamir}
Y. Shamir, Phys. Rev. D71:034509 (2005) [hep-lat/0412014].

\bibitem{Giedt}
J. Giedt, hep-lat/0507002.

\bibitem{Sharpe}
S. R. Sharpe and R. S. van der Water, Phys. Rev. D71:114505 [hep-lat/0409018].

\bibitem{Durr}
S. D\"urr, C. Hoelbling and U. Wenger, Phys. Rev. D70, 094502 (2004) [hep-lat/0406027];
hep-lat/0409108;
S. D\"urr and C. Hoelbling, Phys. Rev. D69, 034503 (2004) [hep-lat/0311002];
hep-lat/0408039;
hep-lat/0411022.

\bibitem{Follana}
E. Follana, A. Hart and C.T.H. Davies [HPQCD Collaboration], 
Phys. Rev. Lett 93:241601 (2004) [hep-lat/0406010];
E. Follana, A. Hart, C.T.H. Davies and Q. Mason, hep-lat/0507011.

\bibitem{Woloshyn}
K. Y. Wong and R. M. Woloshyn, Phys. Rev D71:094508 (2005) [hep-lat/0412001].

\bibitem{Kluberg}
F. Gliozzi, Nucl. Phys. B204, 419 (1982);
A. Duncan, R. Roskies and H. Vaidya, Phys. Lett. B114, 439 (1982);
H. Kluberg-Stern, A. Morel, O. Napoly and B. Petersson, Nucl. Phys. B220, 447 (1983).

\bibitem{Hasenfratz-DeGrand}
A. Hasenfratz and T.A. DeGrand, Phys. Rev. D49, 466 (1994) [hep-lat/9304001].

\bibitem{Smit}
N. Kawamoto and J. Smit, Nucl. Phys. B192, 100 (1981).

\bibitem{Rossi-Wolff}
P. Rossi and U. Wolff, Phys. Rev. D30, 2233 (1984)

\bibitem{Fujikawa}
K. Fujikawa and M. Ishibashi, Nucl. Phys. B605, 365 (2001) [hep-lat/0102012].

\bibitem{Neu1}
H. Neuberger, Phys. Lett. B417, 141 (1998) [hep-lat/9707022].

\bibitem{Bicudo}
D.H. Adams and P. Bicudo, in preparation.

\bibitem{Weiss}
H.S. Sharatchandra, H.J. Thun and P. Weisz, Nucl. Phys. B192, 205 (1981).

\bibitem{Mitra}
P. Mitra and P. Weisz, Phys. Lett. B126, 355 (1983). 

\bibitem{Golterman}
M.F.L. Golterman and J. Smit, Nucl. Phys. B245, 61 (1984).

\bibitem{Luscher(PLB)}
M. L\"uscher, Phys. Lett. B428, 342 (1998) [hep-lat/9802011].

\bibitem{GW}
P.H. Ginsparg and K.G. Wilson, Phys. Rev. D25, 2649 (1982).

\bibitem{Laliena}
P. Hasenfratz, V. Laliena and F. Niedermayer, Phys. Lett. B427, 125 (1998)
[hep-lat/9801021].

\bibitem{Neu2}
H. Neuberger, Phys. Lett. B427, 353 (1998) [hep-lat/9801031].

\bibitem{Fuji(GW)}
K. Fujikawa, Nucl. Phys. B589, 487 (2000) [hep-lat/0004012].

\bibitem{DA(prep)}
D.H. Adams, {\em Perturbative analysis of general lattice fermion determinants},
in preparation.

\bibitem{Duncan}
A. Duncan, E. Eichten, R. Roskies and H. Thacker, Phys. Rev. D60, 054505 (1999)
[hep-lat/9902015].

\bibitem{Duncan1}
A. Duncan, E. Eichten and H. Thacker, Phys. Rev. D59, 014505 (1999)
[hep-lat/9806020].

\bibitem{Shamir(private)}
Y. Shamir, private correspondence.

\bibitem{Wolff}
P. Rossi and U. Wolff, Nucl. Phys. B248, 105 (1984).

\bibitem{Sch}
D.H. Adams and S. Chandrasekharan, Nucl. Phys. B662, 220 (2003) [hep-lat/0303003].

\bibitem{Creutz}
M. Creutz, presentation during the KITP program ``Modern Challenges for Lattice Field 
Theory'' (Jan. 2005); informal discussion at Woody's BBQ, Golita.



\end{thebibliography}
\end{document}